# Machine Learning for Novel Thermal-Materials Discovery: Early Successes, Opportunities, and Challenges

Hang Zhang[1,2], Kedar Hippalgaonkar[3]*, Tonio Buonassisi[4]*, Ole M. Løvvik[5,6]*, Espen Sagvolden[5], Ding Ding[7]*

1 Institute of Engineering Thermophysics, Chinese Academy of Sciences, Beijing 100190, PR China
2 University of Chinese Academy of Sciences, Beijing 100049, PR China
3 Institute of Materials Research and Engineering, A*STAR Research Entities, 138634 Singapore
4 Massachusetts Institute of Technology, Cambridge 02139, MA, USA
5 SINTEF Materials Physics, 0314 Oslo, Norway
6 Department of Physics, University of Oslo, 0316 Oslo, Norway
7 Singapore Institute of Manufacturing Technology, A*STAR Research Entities, 138634 Singapore

*E-mail: kedarh@imre.a-star.edu.sg (K.H.); buonassisi@mac.com (T.B.); olemartin.lovvik@sintef.no (O.L.); ding_ding@simtech.a-star.edu.sg (D.D.)

Abstract

High-throughput computational and experimental design of materials aided by machine learning have become an increasingly important field in material science. This area of research has emerged in leaps and bounds in the thermal sciences, in part due to the advances in computational and experimental methods in obtaining thermal properties of materials. In this paper, we provide a current overview of some of the recent work and highlight the challenges and opportunities that are ahead of us in this field. In particular, we focus on the use of machine learning and high-throughput methods for screening of thermal conductivity for compounds, composites and alloys as well as interfacial thermal conductance. These new tools have brought about a feedback mechanism for understanding new correlations and identifying new descriptors, speeding up the discovery of novel thermal functional materials.

## *Introduction*

As humanity's energy demand increases, so does the demand on materials' thermal and thermal-transport properties. For example, materials with thermal conductivity ($\kappa$) below 1 W/mK are needed for thermal insulation[1,2], and above 10,000 W/mK for heat management of next-generation consumer electronics and energy-generation technologies [3,4]. To gain market acceptance, new materials must also satisfy other application-specific constraints, including electrical conductivity, cost, density, manufacturability, mechanical properties, durability, chemical compatibility, and environmental impact [5]. From a scientific point of view, thermal and thermal-transport properties are governed by a handful of underlying materials properties factoring into the Boltzmann Transport Equation (BTE) that depend on structure and atomic constitution [6]. In essence, the materials-innovation challenge in thermal sciences is more complex than multi-parameter optimization, as it involves not just the search for new materials, but also the search for new physics.

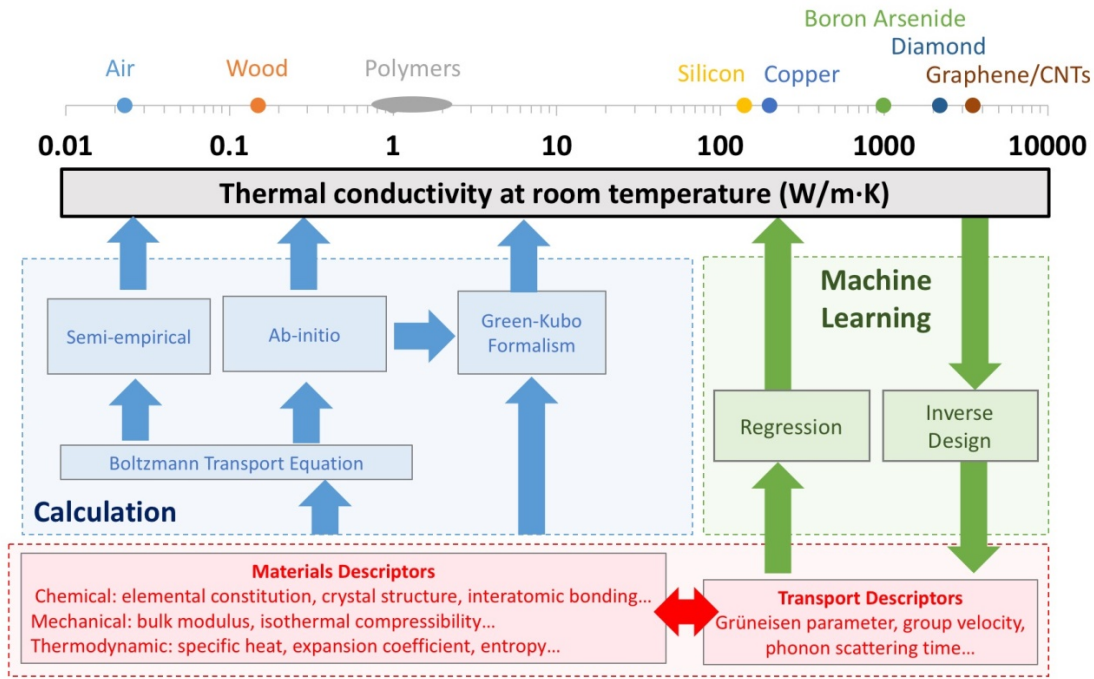

Fig. 1. The range of thermal conductivity can span more than six orders of magnitude. Obtaining accurate values of thermal conductivity and understanding the physical reason for the thermal conductivity is a challenge as calculations are computationally expensive and is typically a one-directional model-based approach of tweaking and observing. With machine learning, the cycle of materials discovery is now complete where we can correlate thermal conductivity with descriptors, providing the feedback to speed up materials innovation and discovery.

Increasingly, materials science researchers are applying combinations of emergent machine learning (ML), high-performance computing (HPC), and automation tools to accelerate the rate of novel materials discovery and development [7]. This transformation in how we perform R&D reflects the community's desires for faster cycles of learning, deeper physical insights, greater sophistication in how we design, synthesize, and optimize materials, and a recognition that we must push physical limits if we are to make meaningful advances in this field of market- and societally-relevant timeframes. This review focuses on the thermal sciences, and is divided into two parts. First, we assess the state of the art in applying machine-learning methods to accelerate materials development. Second, we review current challenges attracting researchers' attention, as well as under-served areas in machine-learning methods for thermal sciences. In the outlook section, we describe future work standing between our present reality and a future vision of fully-automated, self-driving laboratories that enable accelerated materials discovery and development.

As shown in Fig. 1, early successes focus on high-accuracy yet computationally expensive HPC methods to calculate underlying materials properties governing thermal transport. These parameters, which include phonon density of states [8], thermal conductivity[6], Debye temperature[9], and the elastic properties[10-14], can now be routinely computed using variants of density functional theory (DFT)[15, 16] with reasonable accuracy [6], albeit at a rate of a few dozens of compounds per year. To increase throughput and enable screening on the scale of hundreds of thousands of compounds, heuristic models and numerical approximations have been developed [17-20], however with limited accuracies typically in the range of 20%, and with inability to go beyond interpolation. Additionally, new parameters of merit, for example the large splitting of acoustic and optical phonon branches exemplified in boron arsenide [21], are being reported at a rate of a handful per decade. To experimentally validate these predictions, consensus has emerged surrounding best practices for thermal property and transport characterization[22], which eliminate most experimental artifacts and establish community-wide benchmarking [23].

Looking ahead, one important challenge is to develop faster and more accurate predictors of materials descriptors, toward enabling materials searches including millions of compounds. Machine learning has proven useful in this and related domains, to accelerate, augment, and even leapfrog DFT, revealing the difficult-to-calculate parameter $\kappa$ [24, 25] (Fig. 1). Ultimately, extending these predictive tools beyond interpolation may enable new physics-based descriptors to be discovered. High-throughput synthesis tools hold promise to solve the multi-parameter optimization challenge intrinsic to the thermal sciences and related energy fields, as well as to provide valuable feedback to refine theoretical models. Machine-learning methods will be challenged by the unique topology of the thermal sciences, including the high degree of correlation between parameters influencing the BTE, complexity (*e.g.*, microstructures and composition) across multiple length scales, and sparse — but extremely rich — data sets. Data challenges include transferring learnings across different platforms, and integrating disparate data repositories. Physics-based challenges include exploring wave effects of phonons, exceeding the amorphous (Cahill-Pohl) limit[26] on the low end of thermal conductivity, extending phonon scattering times on the high end, and understanding new phenomena such as hydrodynamic scattering of phonons [27].

The thermal properties of materials are very important for understanding thermodynamic stability of structural phases and their suitability for a variety of applications. High thermal conductivity materials are essential for efficient heat removal while low thermal conductivity materials can give rise to the next generation of thermoelectric materials and thermal barriers. Within computational abilities and experimental fabrication and testing, our community has successfully predicted [21], synthesized and experimentally measured the highest thermal conductivity material know to date [28-30]. At the same time, computational predictions and experimental verification have also led to discovery of a number of low thermal conductivity materials [31-35]. While systematic studies of classes of materials to guide us into understanding what material attributes contribute towards thermal conductivity have been undertaken, an approach that scales to the large number of hitherto undiscovered compounds is still lacking.

To do so, we will require a high throughput platform whereby computational screening and experimental testing are conducted on a large number of samples within reasonable computational and experimental resources [7]. High-throughput (HT) computational screening is a rapidly expanding area of materials research[36]. Increasing availability of computational resources have resulted in large databases, and has generated, for example, the AFLOWLIB.org consortium[37], the Materials Project database[36, 38, 39], Citrination[40], among others[41]. These databases and the application of HT methods have recently led to new insights and novel compounds in different fields [41-46]. However, despite the importance of thermal transport properties for many crucial technologies, there are to date only a few HT investigations into lattice thermal conductivity [17, 24, 25, 47, 48].

One main issue is that the determination of the thermal conductivity of materials is computationally demanding as it requires calculation of multiple-phonon scattering processes. A brief overview of the calculation methods has been summarized in Fig. 1. The third-order anharmonic inter-atomic force constants (IFCs) required in order to account for three-phonon scattering processes [6] with standard ways such as density functional theory (DFT) and density functional perturbation theory (DFPT) [49] are generally computationally expensive. This can either be based on the frozen-phonon approach [50] or the temperature-dependent effective potential (TDEP) method [51], the latter based on first-principles molecular dynamics calculations at explicit temperatures, with the option to generate a canonical ensemble of supercell configurations using Monte Carlo sampling.

An alternative approach to calculating thermal conductivity is based on the Green-Kubo formulation which employs molecular dynamics simulations to calculate heat fluxes upon thermal equilibrium [52, 53]. This technique accounts for high-order scattering processes, but semi empirical potentials used in these calculations can lead to errors on the order of 50% [54]. A variety of simple methods have been developed to obtain the thermal properties of materials at reduced computational cost. Early implementations to compute the lattice thermal conductivity were based on semi-empirical models to solve the BTE with some parameters obtained from fitting to experimental data [18, 19]. This reduces the predictive power of semi-empirical methods. Overall, the methods described above are unsuitable for HT generation and screening of large databases of materials properties in order to identify trends and simple descriptors for thermal properties [7], which is where machine learning approaches have demonstrated immense potentials.

## *Early successes – theoretical modeling of thermal properties*
### a. *Bulk stoichiometric compounds*

Despite the computational cost, there have been a few attempts to generate HT data for predicting thermal properties for bulk stoichiometric compounds. The first large scale HT attempt was done by Carrete et al. [24] The work concentrated on half-Heusler (HH) compounds where unstable and zero-gapped compounds were screened by DFT. Then, full DFT calculations for lattice thermal conductivity $\kappa$ for a smaller subset of such compounds were carried out as a training set upon which a ML method known as random forest regression is used to build a classification model for the descriptors. IFCs were also predicted from random forest algorithm upon which good agreement has been obtained. Figure 2(a) shows the distribution lattice parameter of $\kappa$ for low and high thermal conductivity. The group of materials with low thermal conductivity tends to have larger lattice parameters. In fact, the work found that compounds are most likely to have low thermal conductivity if the average atomic radius of the atoms in structural positions are large [24]. HT screening based on another ML approach known as Bayesian optimization was applied to a few classes of compounds by Seko et al. [25]. Figures 2(b) and (c) show two descriptors, namely volume and density, that have been discovered to correlate with $\kappa$.

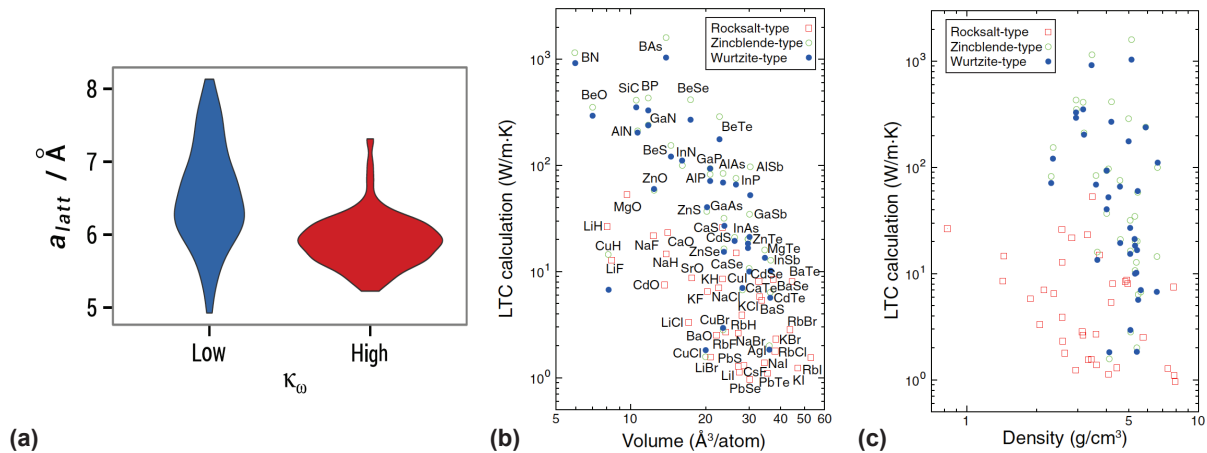

Fig. 2. (a) Distribution of lattice parameter $a_{latt}$ for low and high classification of lattice thermal conductivity obtained from random-forest regression [24]. Larger atomic radii of the structural positions tend to result in lower thermal conductivity. (b,c) Lattice thermal conductivity calculated from first-principles along with their (b) volume and (c) density, respectively [25]. (b) and (c) have been reproduced with permission from the American Physics Society.

Despite the increased speed with such ML methods, a training dataset is still required upon which full calculations are necessary. There exist parallel efforts to develop computationally less expensive ways of calculating $\kappa$. Original models such as semi-empirical methods by Allen [18] and Callaway [19] usually require fitting parameters from experimental data. A recent attempt by Miller et al. (Fig. 3a) [17] uses DFT data to fit to the semi-empirical Debye-Callaway model [19]. This work improved the previous model by incorporating the dependence of coordination number in the Grüneisen parameter. Yet most HT approaches still rely on DFT to some extent. For instance, a much less computationally expensive approach called the Automated GIBBS Library (AGL) [48] has used a quasi-harmonic Debye model [9] to enable a fast HT method for computing $\kappa$ for a large class of materials. This approach only requires electronic DFT calculations to estimate the Grüneisen parameter and is comparable to the accuracy of ML methods (see Fig. 3b). At the same time, principal component analysis has been used to extrapolate IFCs at finite temperatures from a few sets of full IFC calculations to predict the thermal stability and $\kappa$ at finite temperatures [47]. Qin and Hu [55] described a way based on the analysis of the harmonic (second order) IFCs to accelerate the evaluation process of obtaining accurate $\kappa$ by solving the cutoff distance problem. More recently, efforts have been devoted to evaluate the phonon band structures and evaluation of thermodynamic properties for a large number of compounds [56]. For instance, Atsushi Togo's phonon database (http://phonondb.mtl.kyoto-u.ac.jp) with full phonon band structures and derived quantities for 1521 semiconducting inorganic crystal were recently reported (Fig. 3c) [8]. With further development in HT methods for IFCs, we foresee that truly HT computational screening of $\kappa$ with DFT is possible in the near future.

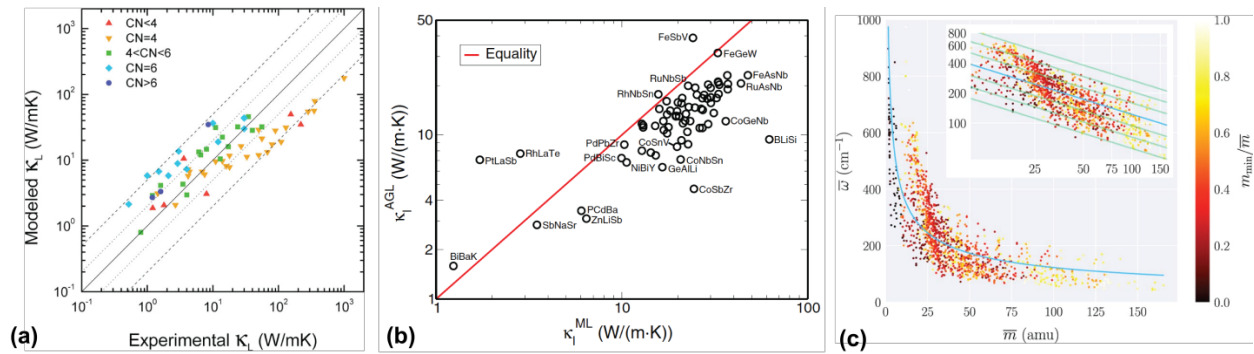

Fig. 3. (a) Semi-empirical model of $\kappa$ that leads to a better agreement with experimental data through incorporation of coordination number (CN). (b) Thermal conductivities of half-Heusler semiconductors at 300 K predicted from the from Automated GIBBS Library versus machine-learning predictions from Carette et al. [24] (c) Large data sets on phonon properties allow for correlations such as the average phonon frequency versus the average atomic mass with data to be obtained [8]. The inset shows a zoom of the data in a log-log scale. The blue line represents a hyperbolic fit of the data, while green lines indicate hypothetical hyperbolic behavior with different constants. Figures (a) and (b) have been reproduced with permission from the American Chemistry Society [17] and the American Physics Society [48], respectively.

### b. *Aperiodic composites and porous materials*

Unlike periodic compounds, composite materials and porous media have wide engineering applications but their effective thermal conductivity is a problem that first principle atomistic methods typically cannot tackle. To predict the effective thermal conductivities of composite materials, existing methods such as effective medium theory (EMT) [57], heat diffusion equation [58], and Boltzmann transport equation (BTE) [59] have been used. The EMT provides an analytical model that can estimate the effective thermal conductivities of the composite materials but its accuracy is limited as it does not account for the effect of heterogeneous distribution of constituent materials. In order to take in to account the details of materials distribution in a composite, direct solutions of heat diffusion equation will be necessary. Many numerical methods such as finite volume method (FVM) [60], the finite element method (FEM) [58] and the lattice Boltzmann method (LBM) [61] have also been developed. All these approaches are based on solving partial differential equations (PDE) which are computationally costly. Recently, Zhang et al. [62] designed a genetic algorithm to optimize the configuration of silicon-germanium composites under multi-parameters for the best thermal conductivities (Fig. 4). Possessing the advantages of low cost for both coding and computational expenses, this approach can be feasibly grafted for solving optimization problems on thermal properties of other composites.

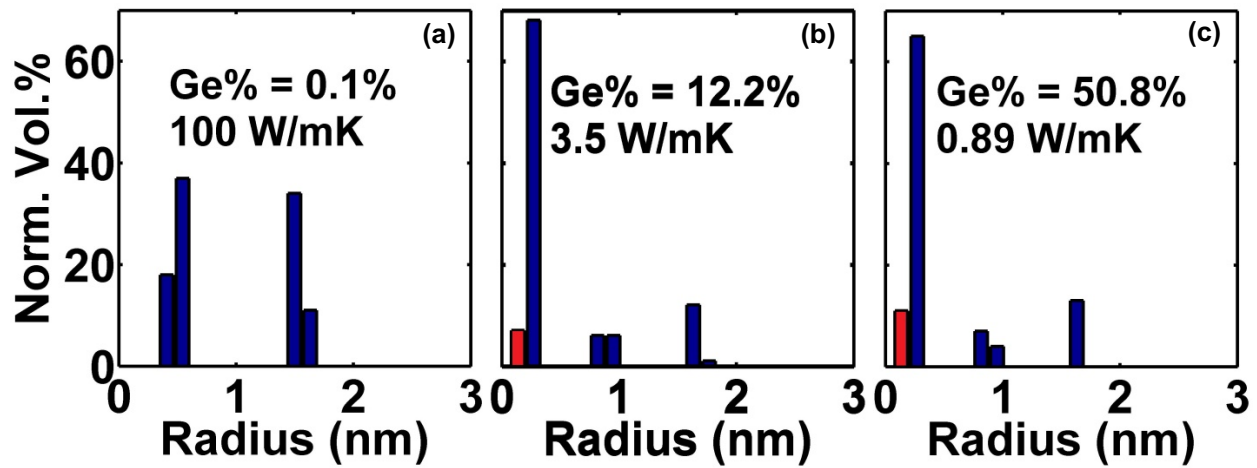

Fig. 4. Optimal size distribution of Ge nanoparticles for lowest thermal conductivity of $Si_xGe_{1-x}$ composites. Heights of histogram bars indicate fraction of Ge allocated to each nanoparticle size. The best distribution introduces additional non-adjacent peaks as more Ge is added. Red histogram bars indicate point defects [62].

In another work by Wei et al. [63], machine learning methods including Support Vector Regression (SVR), Gaussian Process Regression (GPR) and Convoluted Neural Networks (CNN) were used to study the effective thermal conductivity of composite materials or porous media. SVR and GPR are non-linear regression methods that provide thermal conductivity prediction with certain pre-defined descriptors, while CNN can directly extract structural features from figures (as descriptors) and then predict the thermal conductivity. Comparing to EMT, which is based on physical understanding of the system, machine learning is based solely on data analysis. Wei et al. created a database using the quartet structure generation set (QSGS) to generate composite material structure and applied LBM to calculate the effective thermal conductivity. Then, this database was used to train and test the different machine learning methods. SVR and GPR have shown good predictions at a fraction of the computational cost. CNN is the most comprehensive in extracting geometric features but it is only accurate with larger training datasets.

### c. *Alloys – using High Entropy Alloys as a test case*

A particularly interesting group of materials exhibiting exotic properties and ultra-low thermal conductivity is formed when mixing several elements on an atomic level with random occupation of each lattice site, i.e. solid solution. However, it is challenging to describe such materials from first principles. Since there is perfect disorder in which atom occupies which position, there is no longer a well-defined translational symmetry. A particularly interesting test-case of such multi-principal element alloys are high-entropy alloys (HEAs) when they are based on different elements and constitute a single phase with solid solution [64]. Alloying is a well-known technique to reduce the thermal conductivity, and the HEA concept has therefore been used to minimize thermal conductivity in several studies, often with an emphasis to develop novel thermoelectric materials [65].

Several methods have been employed to describe HEAs on the electronic scale [66], including the virtual crystal approximation (VCA) [12], coherent potential approximation (CPA) [67], special quasi-random structures (SQS) [14, 68-72], and molecular dynamics (MD) based methods [73-75]. Only a few studies have assessed the thermal conductivity of HEAs, using semi-empirical MD [76, 77]. As mentioned above, the error involved in such methods may be quite high. This is particularly so for complex compounds with a large number of pair- and higher-order potentials. The studies only describe qualitative features of heat transport in generalized HEAs and are thus not suited for HT investigations. The VCA method has been used to predict thermal conductivity of solid solution alloys. It is based on first principles and requires relatively expensive calculations of interatomic force constants within a phonon scheme. Nevertheless, interpolation of force constants makes this method rather efficient, and has been shown to reproduce experimental data relatively well in e.g. the entire ternary phase diagram of the solid solution (Ti,Zr,Hf)NiSn system (Fig. 5a) [78]. In order to obtain reliable and predictive results, however, the most applicable of the above methods to computing thermal transport appears to be the SQS construction. In this method, the atomic positions of a supercell of finite size are designed to ideally mimic the nuclear pair-correlation function of the solid solution. The larger the cell, the better the pair-correlation function can be approximated. The technique gives excellent results; 128-atom supercells were recently found to reproduce very well the experimental thermal conductivity of the random alloy $In_{1-x}Ga_xAs$ using SQS with a Green's function approach [79]. It was shown in the same study that disorder of the interatomic force constants was necessary to obtain good correspondence with experiment, rendering the VCA approach significantly less accurate (Fig. 5b).

So far, no studies in the literature have, to the best of our knowledge, predicted the calculated thermal conductivity of a HEA based on SQS at the DFT level. It can be anticipated that such studies will be available quite soon, and that they will form the basis of HT studies aiming at developing HEAs with extremely low thermal conductivity. Since these materials often come with other extraordinary properties, accelerated discoveries of novel HEAs will lead to materials with unique combinations of thermal and other properties.

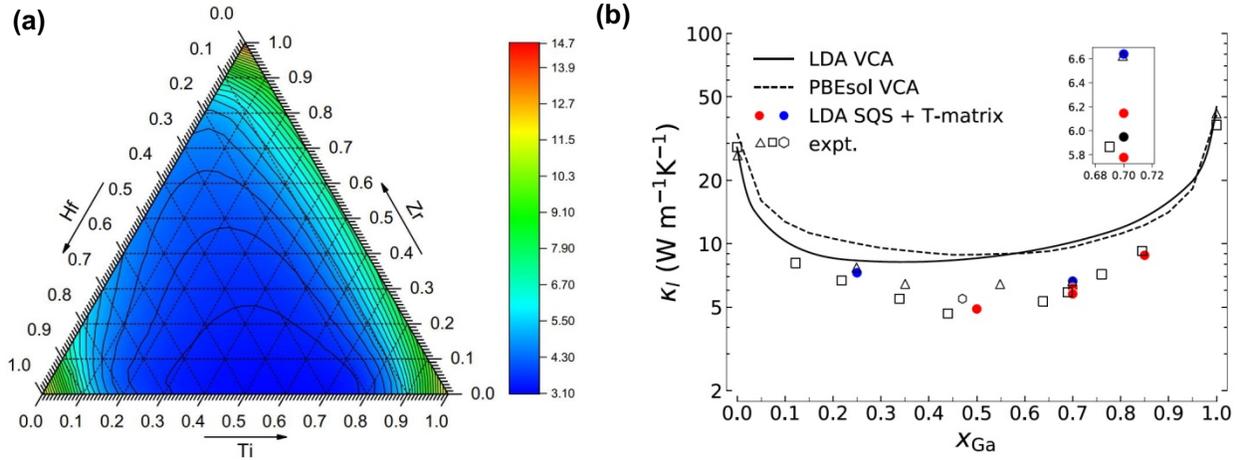

Fig. 5. The phonon part of the thermal conductivity calculated with DFT employing semi classical Boltzmann transport equations. (a) $Ti_xZr_yHf_{1-x-y}NiSn$ at 300 K using the virtual crystal approximation (VCA). The bottom right corner corresponds to TiNiSn, the top to ZrNiSn, and the bottom left to HfNiSn. (b) $In_{1-x}Ga_xAs$ at 300 K using VCA (solid line based on the local density approximation (LDA) and dashed line based on the PBEsol generalized gradient approximation) and the special quasi-random structure (SQS) technique (red and blue dots based on LDA with 128- and 250-atom supercells, black dot based on PBEsol). Experimental results are shown with the open symbols. The figures have been reproduced with permission from the American Physics Society[78, 79].

### d. *Interfacial thermal conductance – beyond the Acoustic and Diffuse Mismatch models*

Ever since Kapitza[80] discovered a non-continuous temperature drop at the interface between helium and a solid, interfacial thermal conductance has become a central problem to thermal material design. The interfacial thermal conductance between two materials is the ratio of the temperature discontinuity at the interface to the power per unit area flowing across the interface. The Acoustic Mismatch Model (AMM) and the Diffuse Mismatch Model (DMM) [81, 82] provide the upper and lower bounds for such an estimate, assuming no scattering for the former and completely diffusive scattering for the latter. Such models have proved useful for solid-solid interfaces and provided significant insight into the mechanics of phonon transport at the interface – the group velocity of the phonons and the overlap of the phonon density of states being two key physical descriptors that are responsible for how heat flows at an interface. However, it is not accurate for describing interfaces with real defects and roughness[82]. A significant advancement was made due to full calculations of the phonon density of states and integrating over this in the Boltzmann transport simulations or Green-Kubo (numerical or analytical integration techniques) [83] allows for the description of frequency dependent interfacial conductance. However, such methods can lead to a loss of accuracy by discounting effects such as intermixing at the interfaces, roughness effects and electron-phonon coupling, and not properly accounting for the finite size effects of the simulation domain [84]. While lots of work have shown the limited validity of AMM and DMM [85, 86] and new models have been proposed [87, 88], ML methods can certainly bridge this gap in knowledge, by identifying the key physical attributes necessary to accurately predict the thermal interface properties.

Two recent works have utilized ML methods to understand interfacial thermal conductance. One work[89] explored the use of Bayesian optimization for computing interfacial thermal conductance (ITC) of super lattice structures consisting of Si and Ge (Fig. 6a). ITC can be minimized or maximized by combining Atomistic Green's Function (AGF) with Bayesian Optimization just by calculating only a few percent of all possible structures, leading to considerable savings in computational resources. It is also found that aperiodic structures can minimize ITC more due to a lack of phonon coherence. More recently, data-driven approaches[90] utilize online and published data with AMM and DMM. Using ML regression methods including SVR, GVR, accurate prediction of interfacial thermal conductance has been achieved. It is found that AMM and DMM are not good descriptors while melting point and heat capacity are good descriptors (Fig. 6b). Even more recently, ML based regression and CNN have recently been employed to study interfacial thermal conductance between graphene and hexagonal boron nitride [91].

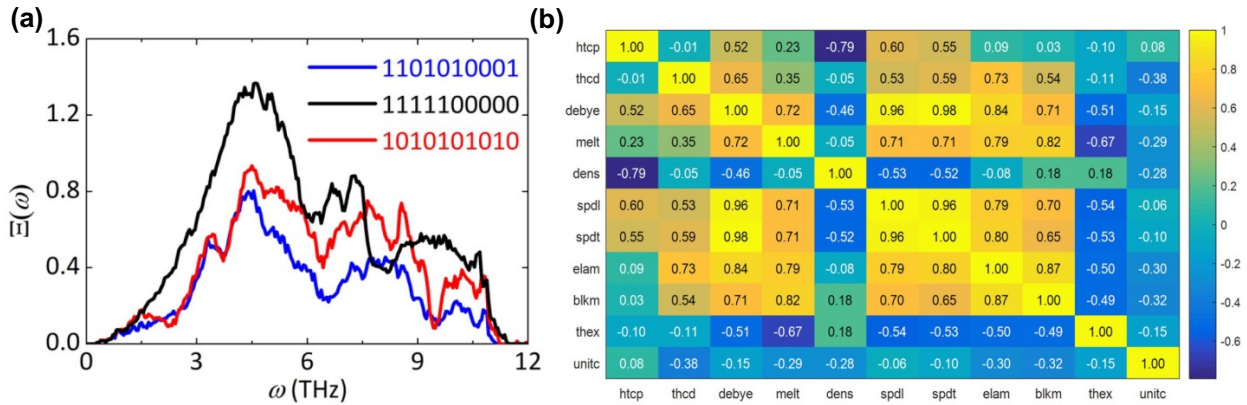

Fig. 6. (a) The phonon transmission for three superlattice structures optimal aperiodic superlattice (1101010001), and periodic superlattice with the largest (1111100000) and smallest (1010101010) periodic thickness, where, "1" and "0" indicate the unit layer consists of Ge and Si, respectively [89]. The aperiodic structure has the lowest transmission across all phonon frequencies. (b) Pearson correlation coefficient map between different materials properties [90]. htcp (heat capacity), thcd (thermal conductivity), debye (Debye temperature), melt (melting point), dens (density), spdl (speed of sound longitudinal), spdt (speed of sound transverse), elam (elastic modulus),

blkm (bulk modulus), thex (thermal expansion coefficient), and unitc (unit cell volume). For example, debye and spdl (spdt) are strongly positively correlated.

*Experiments*

While HT predictions of $\kappa$ through ML have been explored, experimental efforts to realize this realm have been lacking. In principle, techniques such as time domain thermoreflectance [92] (TDTR), the broadband frequency domain thermoreflectance [93] (FDTR) and other nanoscale scanning techniques [94] are perfectly suited for fast, non-destructive testing of thermal properties. Implemented with sophisticated modeling based on the approaches described above, these high throughput experimental tools can ascertain the thermal conductivity, specific heat (extracted from the thermal diffusivity), interfacial thermal conductance as well as anisotropic thermal properties. Going beyond, because of the depth of knowledge developed by using frequency-dependent intrinsic transport descriptors for phonon mean-free paths[95], interfacial thermal conductance [96], surface roughness [97, 98], phonon dispersion[99] etc. combined with high-throughput materials synthesis techniques[100, 101] both in the solid solution and combinatorial thin film form, there is a huge opportunity where such HT experimental techniques can provide high-fidelity data that will not only provide insight on what dictates thermal transport, but also provide a database that can serve as a test-set for ML algorithms. Similar approaches have already proven successful for small molecules and proteins[102, 103] and the need-of-the-hour is to leverage upon fast experimental characterization tools to create a thermal material property library. For example, measuring the in-plane and cross-plane thermal conductivities of a material grown by combinatorial synthesis could go a long way towards designing new thermal materials. First steps in this direction utilize time domain thermoreflectance to measure $\kappa$ experimentally for nickel solid solutions [104, 105]. Such HT techniques have already been being developed for chemistry[106], solar cell materials[101] and batteries[107].

Sophisticated ML tools such as Bayesian inference are able to use a forward model (similar to the Boltzmann Transport Equation) in order to extract hidden materials and transport properties simultaneously. As an example, by looking at a system-level model in a photovoltaic cell, one is able to extract the bulk and interface properties that are limiting its performance [44]. Similar new hardware approaches can also be developed for thermoelectric materials. One can envision utilizing such ML algorithms coupled with advanced statistical analysis to provide experimental tools (for example, extending the theoretical models used with TDTR) to provide fast screening for measurement of thermal properties of a large class of compounds. Large scale synthesis of bulk and thin film materials with varying stoichiometry, microstructure, dopant ratios and physical properties is an outstanding challenge that will impact not just the thermal community but other fields of materials science as well. In addition, a clear and present opportunity is to leverage upon existing theoretical databases and integrate them into a common language that is widely available to researchers working in this and related areas. This will enable an artful application of ML to sparse, but high-quality datasets (experimentally generated, but amplified by theoretical calculations). The general approach here is to provide high-throughput experimental data embellished by fast theoretical predictions, while high-fidelity careful measurements can subsequently be performed after this initial screening process. Since little work has been performed in this area to this date, there is a vast space for discovery of new materials with novel thermal properties and exhibiting new physics; for example, moving beyond the classical size effects and scattering of phonons as particles, towards wave and coherent effects[108]. Such high-throughput synthesis and thermal characterization will also enable the development of holistic understanding of thermal properties on material classes, bonding, alloys, microstructures, defects (both from understanding and moving towards engineering).

## *Conclusions and Outlook*

Thanks to the establishment of materials database frameworks and the rapid development on both computational hardware and algorithms for machine learning (ML), pioneering works have emerged in this interdisciplinary field of data science and materials discovery for thermal applications. While algorithms for data processing and ML have become increasing sophisticated, high-quality datasets particularly suitable for thermal properties are still difficult to obtain. This is largely due to the high computational cost involved in computing various thermal properties from physical models and first-principles. Therefore, development of a high-throughput methodology has been one of the most important challenges, which can clear the bottleneck of data deficiency. The second frontier is our lack of comprehensive understanding on the correlation of material descriptors with transport properties. ML has provided us new insights into correlations that were not physically intuitive, offering us insights into future material discoveries for thermal science. Design of new high-throughput (HT) experiments enabled by ML will play a key role in augmenting datasets, but only if the community provides these on open data platforms that are accessible to all practitioners. Overall, the coupling of theoretical and experimental HT techniques is a vital tool in the development of this field, liberating us from the repetitive work of parameter sweeps and measurements towards new physics and new materials. The confluence of these powerful new approaches along with deep domain expertise will surely take the field in unheralded directions.

## *Conflict of interest*
There are no conflicts to declare.

## *Acknowledgements*
H.Z. acknowledges the financial support from CAS Pioneer Hundred Talents Program. O.L. and E.S. acknowledge the financial support from Research Council of Norway. D.D and K.H. acknowledge support from A*Star's Science and Engineering Research Council, project: A1898b0043 on Accelerating Materials Development for Manufacturing.

## **References**

1. T. Li, J. Song, X. Zhao, Z. Yang, G. Pastel, S. Xu, C. Jia, J. Dai, C. Chen, A. Gong, F. Jiang, Y. Yao, T. Fan, B. Yang, L. Wågberg, R. Yang and L. Hu, *Sci Adv.*, 2018, **4**, eaar3724.
2. B. P. Jelle, *Energy Build.*, 2011, **43**, 2549-2563.
3. M. Behnia, L. Maguire and G. Morrison, 5th International Conference on Thermal and Mechanical Simulation and Experiments in Microelectronics and Microsystems, 2004. EuroSimE 2004. Proceedings of the, 2004.
4. A. L. Moore and L. Shi, *Mater. Today*, 2014, **17**, 163-174.
5. M. F. Ashby, in *Materials and the Environment (Second Edition)*, ed. M. F. Ashby, Butterworth-Heinemann, Boston, 2013, pp. 227-273.
6. D. A. Broido, M. Malorny, G. Birner, N. Mingo and D. A. Stewart, *Appl. Phys. Lett.*, 2007, **91**, 231922.
7. S. Curtarolo, G. L. W. Hart, M. B. Nardelli, N. Mingo, S. Sanvito and O. Levy, *Nat. Mater.*, 2013, **12**, 191-201.
8. G. Petretto, S. Dwaraknath, H. P. C. Miranda, D. Winston, M. Giantomassi, M. J. v. Setten, X. Gonze, K. A. Persson, G. Hautier and G.-M. Rignanese, *Sci. Data*, 2018, **5**, 180065.
9. M. A. Blanco, E. Francisco and V. Luaña, *Comput. Phys. Commun.*, 2004, **158**, 57-72.
10. P. Ravindran, L. Fast, P. A. Korzhavyi, B. Johansson, J. Wills and O. Eriksson, *J. Appl. Phys.*, 1998, **84**, 4891-4904.
11. M. de Jong, W. Chen, T. Angsten, A. Jain, R. Notestine, A. Gamst, M. Sluiter, C. Krishna Ande, S. van der Zwaag, J. J. Plata, C. Toher, S. Curtarolo, G. Ceder, K. A. Persson and M. Asta, *Sci. Data*, 2015, **2**, 150009.
12. M. Q. Liao, Y. Liu, L. J. Min, Z. H. Lai, T. Y. Han, D. N. Yang and J. C. Zhu, *Intermetallics*, 2018, **101**, 152-164.
13. J. A. Rogers, M. Fuchs, M. J. Banet, J. B. Hanselman, R. Logan and K. A. Nelson, *Appl. Phys. Lett.*, 1997, **71**, 225-227.
14. Z. Q. Wen, Y. H. Zhao, J. Z. Tian, S. Wang, Q. W. Guo and H. Hou, *J. Mater. Sci.*, 2019, **54**, 2566-2576.
15. Y. Wang, S.-L. Shang, H. Fang, Z.-K. Liu and L.-Q. Chen, *npj Comput. Mater.*, 2016, **2**, 16006.
16. H. Bao, J. Chen, X. Gu and B. Cao, *ES Energy Environ.*, 2018, DOI: 10.30919/esee8c149.
17. S. A. Miller, P. Gorai, B. R. Ortiz, A. Goyal, D. Gao, S. A. Barnett, T. O. Mason, G. J. Snyder, Q. Lv, V. Stevanović and E. S. Toberer, *Chem. Mater.*, 2017, **29**, 2494-2501.
18. P. B. Allen, *Phys. Rev. B*, 2013, **88**, 144302.
19. J. Callaway, *Phys. Rev.*, 1959, **113**, 1046.
20. D. T. Morelli and G. A. Slack, in *High Thermal Conductivity Materials*, eds. S. L. Shindé and J. S. Goela, Springer New York, New York, NY, 2006, pp. 37-68.
21. L. Lindsay, D. A. Broido and T. L. Reinecke, *Phys. Rev. Lett.*, 2013, **111**, 025901.
22. D. Zhao, X. Qian, X. Gu, S. A. Jajja and R. Yang, *J. Electron. Packaging*, 2016, **138**, 040802-040802-040819.
23. S. Rubin and F. F. Gettinger, *Semiconductor measurement technology: Thermal resistance measurements on power transistors*, 1979.

24. J. Carrete, W. Li, N. Mingo, S. Wang and S. Curtarolo, *Phys. Rev. X*, 2014, **4**, 011019.
25. A. Seko, A. Togo, H. Hayashi, K. Tsuda, L. Chaput and I. Tanaka, *Phys. Rev. Lett.*, 2015, **115**, 205901.
26. D. G. Cahill, S. K. Watson and R. O. Pohl, *Phys. Rev. B*, 1992, **46**, 6131-6140.
27. S. Lee, D. Broido, K. Esfarjani and G. Chen, *Nat. Commun.*, 2015, **6**, 6290.
28. J. S. Kang, M. Li, H. Wu, H. Nguyen and Y. Hu, *Science*, 2018, **361**, 575.
29. S. Li, Q. Zheng, Y. Lv, X. Liu, X. Wang, P. Y. Huang, D. G. Cahill and B. Lv, *Science*, 2018, **361**, 579-581.
30. F. Tian, B. Song, X. Chen, N. K. Ravichandran, Y. Lv, K. Chen, S. Sullivan, J. Kim, Y. Zhou, T. H. Liu, M. Goni, Z. Ding, J. Sun, G. A. G. Udalamatta Gamage, H. Sun, H. Ziyaee, S. Huyan, L. Deng, J. Zhou, A. J. Schmidt, S. Chen, C. W. Chu, P. Y. Huang, D. Broido, L. Shi, G. Chen and Z. Ren, *Science*, 2018, **361**, 582-585.
31. Z. Zhang, S. Hu, T. Nakayama, J. Chen and B. Li, *Carbon*, 2018, **139**, 289-298.
32. X. Xu, J. Chen, J. Zhou and B. Li, *Adv. Mater.*, 2018, **30**, 1705544.
33. W. Lee, H. Li, A. B. Wong, D. Zhang, M. Lai, Y. Yu, Q. Kong, E. Lin, J. J. Urban, J. C. Grossman and P. Yang, *PNAS*, 2017, **114**, 8693-8697.
34. L.-D. Zhao, S.-H. Lo, Y. Zhang, H. Sun, G. Tan, C. Uher, C. Wolverton, V. P. Dravid and M. G. Kanatzidis, *Nature*, 2014, **508**, 373-377.
35. C. Chiritescu, D. G. Cahill, N. Nguyen, D. Johnson, A. Bodapati, P. Keblinski and P. Zschack, *Science*, 2007, **315**, 351-353.
36. A. Jain, G. Hautier, C. J. Moore, S. Ping Ong, C. C. Fischer, T. Mueller, K. A. Persson and G. Ceder, *Comput. Mater. Sci.*, 2011, **50**, 2295-2310.
37. S. Curtarolo, W. Setyawan, G. L. W. Hart, M. Jahnatek, R. V. Chepulskii, R. H. Taylor, S. Wang, J. Xue, K. Yang, O. Levy, M. J. Mehl, H. T. Stokes, D. O. Demchenko and D. Morgan, *Comput. Mater. Sci.*, 2012, **58**, 218-226.
38. A. Jain, S. P. Ong, G. Hautier, W. Chen, W. D. Richards, S. Dacek, S. Cholia, D. Gunter, D. Skinner, G. Ceder and K. A. Persson, *APL Mater.*, 2013, **1**, 011002.
39. S. P. Ong, W. D. Richards, A. Jain, G. Hautier, M. Kocher, S. Cholia, D. Gunter, V. L. Chevrier, K. A. Persson and G. Ceder, *Comput. Mater. Sci.*, 2013, **68**, 314-319.
40. J. O'Mara, B. Meredig and K. Michel, *JOM*, 2016, **68**, 2031-2034.
41. J. E. Saal, S. Kirklin, M. Aykol, B. Meredig and C. Wolverton, *JOM*, 2013, **65**, 1501-1509.
42. K. Yang, W. Setyawan, S. Wang, M. Buongiorno Nardelli and S. Curtarolo, *Nat. Mater.*, 2012, **11**, 614-619.
43. G. Ceder, *MRS Bull.*, 2010, **35**, 693-701.
44. R. E. Brandt, R. C. Kurchin, V. Steinmann, D. Kitchaev, C. Roat, S. Levcenco, G. Ceder, T. Unold and T. Buonassisi, *Joule*, 2017, **1**, 843-856.
45. A. Seko, T. Maekawa, K. Tsuda and I. Tanaka, *Phys. Rev. B*, 2014, **89**, 054303.
46. A. N. Kolmogorov, S. Shah, E. R. Margine, A. F. Bialon, T. Hammerschmidt and R. Drautz, *Phys. Rev. Lett.*, 2010, **105**, 217003.
47. A. van Roekeghem, J. Carrete, C. Oses, S. Curtarolo and N. Mingo, *Phys. Rev. X*, 2016, **6**, 041061.

48. C. Toher, J. J. Plata, O. Levy, M. de Jong, M. Asta, M. B. Nardelli and S. Curtarolo, *Phys. Rev. B*, 2014, **90**, 174107.
49. G. Deinzer, G. Birner and D. Strauch, *Phys. Rev. B*, 2003, **67**, 144304.
50. J. M. Skelton, S. C. Parker, A. Togo, I. Tanaka and A. Walsh, *Phys. Rev. B*, 2014, **89**, 205203.
51. O. Hellman and D. A. Broido, *Phys. Rev. B*, 2014, **90**, 134309.
52. M. S. Green, *J. Chem. Phys.*, 1954, **22**, 398-413.
53. R. Kubo, *J. Phys. Soc. Jpn.*, 1957, **12**, 570-586.
54. M. Zebarjadi, K. Esfarjani, M. S. Dresselhaus, Z. F. Ren and G. Chen, *Energy Environ. Sci.*, 2012, **5**, 5147-5162.
55. G. Qin and M. Hu, *npj Comput. Mater.*, 2018, **4**, 3.
56. F. Legrain, J. Carrete, A. van Roekeghem, S. Curtarolo and N. Mingo, *Chem. Mater.*, 2017, **29**, 6220-6227.
57. P. Cheng and C.-T. Hsu, *J. Porous Media*, 1999, **2**, 19-38.
58. Z. Tong, M. Liu and H. Bao, *Int. J. Heat Mass Transfer*, 2016, **100**, 355-361.
59. M. Wang and N. Pan, *Int. J. Heat Mass Transfer*, 2008, **51**, 1325-1331.
60. C. Demuth, M. A. A. Mendes, S. Ray and D. Trimis, *Int. J. Heat Mass Transfer*, 2014, **77**, 979-994.
61. X. He, S. Chen and G. D. Doolen, *J. Comput. Phys.*, 1998, **146**, 282-300.
62. H. Zhang and A. J. Minnich, *Sci. Rep.*, 2015, **5**, 8995.
63. H. Wei, S. Zhao, Q. Rong and H. Bao, *Int. J. Heat Mass Transfer*, 2018, **127**, 908-916.
64. D. B. Miracle and O. N. Senkov, *Acta Mater.*, 2017, **122**, 448-511.
65. W. R. Zhang, P. K. Liaw and Y. Zhang, *Sci. China. Mater.*, 2018, **61**, 2-22.
66. F. Y. Tian, *Front. Mater.*, 2017, **4**.
67. M. Ogura, T. Fukushima, R. Zeller and P. H. Dederichs, *J. Alloys Compd.*, 2017, **715**, 454-459.
68. R. Z. Zhang, F. Gucci, H. Y. Zhu, K. Chen and M. J. Reece, *Inorg. Chem.*, 2018, **57**, 13027-13033.
69. H. Q. Song, F. Y. Tian and D. P. Wang, *J. Alloys Compd.*, 2016, **682**, 773-777.
70. S. Wang, T. Zhang, H. Hou and Y. H. Zhao, *Phys. Status Solidi B*, 2018, **255**.
71. T. T. Zuo, M. C. Gao, L. Z. Ouyang, X. Yang, Y. Q. Cheng, R. Feng, S. Y. Chen, P. K. Liaw, J. A. Hawk and Y. Zhang, *Acta Mater.*, 2017, **130**, 10-18.
72. F. Kormann, Y. Ikeda, B. Grabowski and M. H. F. Sluiter, *npj Comput. Mater.*, 2017, **3**, 9.
73. H. Babaei, P. Keblinski and J. M. Khodadadi, *J. Appl. Phys.*, 2012, **112**.
74. A. Giri, J. L. Braun, C. M. Rost and P. E. Hopkins, *Scr. Mater.*, 2017, **138**, 134-138.
75. S.-W. Kao, J.-W. Yeh and T.-S. Chin, *J. Phys. Condens. Matter*, 2008, **20**.
76. M. Caro, L. K. Beland, G. D. Samolyuk, R. E. Stoller and A. Caro, *J. Alloys Compd.*, 2015, **648**, 408-413.
77. A. Giri, J. L. Braun and P. E. Hopkins, *J. Appl. Phys.*, 2018, **123**.
78. S. N. H. Eliassen, A. Katre, G. K. H. Madsen, C. Persson, O. M. Lovvik and K. Berland, *Phys. Rev. B*, 2017, **95**.
79. M. Arrigoni, J. Carrete, N. Mingo and G. K. H. Madsen, *Phys. Rev. B*, 2018, **98**, 8.

80. in *Collected Papers of P.L. Kapitza*, ed. D. Ter Haar, Pergamon, 1965, pp. 581-624.
81. W. A. Little, *Can. J. Phys.*, 1959, **37**, 334-349.
82. E. T. Swartz and R. O. Pohl, *Rev. Mod. Phys.*, 1989, **61**, 605-668.
83. E. S. Landry and A. J. H. McGaughey, *Phys. Rev. B*, 2009, **80**, 165304.
84. Z. Liang and M. Hu, *J. Appl. Phys.*, 2018, **123**, 191101.
85. T. Zhan, S. Minamoto, Y. Xu, Y. Tanaka and Y. Kagawa, *AIP Adv.*, 2015, **5**, 047102.
86. N. Yang, T. Luo, K. Esfarjani, A. Henry, Z. Tian, J. Shiomi, Y. Chalopin, B. Li and G. Chen, *J. Comput. Theor. Nanosci.*, 2015, **12**, 168-174.
87. K. Gordiz and A. Henry, *eprint arXiv:1407.6410*, 2014, arXiv:1407.6410.
88. Y. Zhang, D. Ma, Y. Zang, X. Wang and N. Yang, *Front. Energy Res.*, 2018, **6**, 48.
89. S. Ju, T. Shiga, L. Feng, Z. Hou, K. Tsuda and J. Shiomi, *Phys. Rev. X*, 2017, **7**, 021024.
90. T. Zhan, L. Fang and Y. Xu, *Sci. Rep.*, 2017, **7**, 7109.
91. H. Yang, Z. Zhang, J. Zhang and X. C. Zeng, *Nanoscale*, 2018, **10**, 19092-19099.
92. K. Kang, Y. K. Koh, C. Chiritescu, X. Zheng and D. G. Cahill, *Rev. Sci. Instrum.*, 2008, **79**, 114901.
93. K. T. Regner, D. P. Sellan, Z. Su, C. H. Amon, A. J. H. McGaughey and J. A. Malen, *Nat. Commun.*, 2013, **4**, 1640.
94. D. G. Cahill, P. V. Braun, G. Chen, D. R. Clarke, S. Fan, K. E. Goodson, P. Keblinski, W. P. King, G. D. Mahan, A. Majumdar, H. J. Maris, S. R. Phillpot, E. Pop and L. Shi, *App. Phys. Rev.*, 2014, **1**, 011305.
95. F. Yang and C. Dames, *Phys. Rev. B*, 2013, **87**, 035437.
96. W. Zhang, T. S. Fisher and N. Mingo, *J. Heat Transfer*, 2006, **129**, 483-491.
97. N. K. Ravichandran, H. Zhang and A. J. Minnich, *Phys. Rev. X*, 2018, **8**, 041004.
98. J. Lim, K. Hippalgaonkar, S. C. Andrews, A. Majumdar and P. Yang, *Nano Lett.*, 2012, **12**, 2475-2482.
99. N. Mingo, *Phys. Rev. B*, 2003, **68**, 113308.
100. S. M. Senkan, *Nature*, 1998, **394**, 350.
101. Z. Long, H. Ren, J. Sun, J. Ouyang and N. Na, *Chem. Commun.*, 2017, **53**, 9914-9917.
102. K. H. Bleicher, H.-J. Böhm, K. Müller and A. I. Alanine, *Nat. Rev. Drug Discovery*, 2003, **2**, 369.
103. S. A. Sundberg, *Curr. Opin. Biotechnol.*, 2000, **11**, 47-53.
104. S. Huxtable, D. G. Cahill, V. Fauconnier, J. O. White and J.-C. Zhao, *Nat. Mater.*, 2004, **3**, 298-301.
105. X. Zheng, D. G. Cahill, P. Krasnochtchekov, R. S. Averback and J. C. Zhao, *Acta Mater.*, 2007, **55**, 5177-5185.
106. M. Shevlin, *ACS Med. Chem. Lett.*, 2017, **8**, 601-607.
107. W. Lu, R. Xiao, J. Yang, H. Li and W. Zhang, *J. Materiomics*, 2017, **3**, 191-201.
108. G. Xie, D. Ding and G. Zhang, *Adv. Phys. X*, 2018, **3**, 1480417.